\documentstyle[12pt]{article}

\begin{document}
\begin{titlepage}
\begin{flushright}
Alberta-Thy-01-98\\
January,1998\\
\end{flushright}
\begin{center}
{\Large\bf $D_s^+\rightarrow \phi {\rho}^{+}$ Decay}\\
\vskip 5mm
E. El aaoud\\
{\em Theoretical Physics Institute and Department of Physics,\\ University of Alberta,
Edmonton, Alberta T6G 2J1, Canada.}\\
\end{center}
\vskip 5mm
\begin{abstract}
Motivated by the experimental measurement of the decay rate, $\Gamma$, and the longitudinal polarization, $P_L$, in the Cabibbo favored decay $D_s^+\rightarrow \phi {\rho}^{+}$, we have  studied theoretical prediction within the context of factorization approximation invoking several form factors models. We were able to obtain agreement with experiment for both $\Gamma$ and $P_L$ by using experimentally measured values of the form factors $A_1^{D_s\phi}(0)$, $A_2^{D_s\phi}(0)$ and $V^{D_s\phi}(0)$  in the semi-leptonic decay $D_s^+\rightarrow \phi {\it l}^{+}\nu_{\it l}$.  We have also included in our calculation the effect of the final state interaction ($fsi$) by working with the partial waves amplitudes $S$, $P$ and $D$. Numerical calculation shows that  the  decay amplitude is dominated by $S$ wave, and that the polarization is sensitive to the interference between $S$ and $D$ waves. The range of the  phase difference $\delta_{SD} = \delta_S - \delta_D$ accommodated by experimental error in $P_L$ is large. \end{abstract}
PACS index: 13.25.Ft, 13.25.-k
\end{titlepage}

\begin{center}
{\bf I. Introduction}
\end{center}

The branching ratio and the longitudinal polarization in ${D}_{s}^+\rightarrow \phi {\rho}^{+}$ have now been measured:

\begin{eqnarray}
B(D_s^+\rightarrow \phi {\rho}^{+}) = (6.7\pm 2.3)\%~ \cite{ref:pdg}  \nonumber  \\
P_L(D_s^+\rightarrow \phi {\rho}^{+}) \equiv {\Gamma}_{L}/\Gamma = (0.370 \pm 0.035 \pm 0.038).~~~ \cite{ref:balic}
\end{eqnarray}

Theoretically, Gourdin et al. \cite{ref:gour} studied the ratio\\
\begin{equation}
R_h \equiv {B(D_s^+\rightarrow \phi {\rho}^{+})}/{B(D_s^+\rightarrow \phi {\pi}^{+})} = 1.86 \pm 0.26 \pm~^{0.29}_{0.40}.~~~  \cite{ref:pa1}
\end{equation}

Within the context of the factorization scheme, which the authors of [3] adopt, this ratio is independent of the normalization of the form factor $A_1(0)$. It depends on the ratios
\begin{equation}
x\equiv A_2(0)/A_1(0), \hspace{.5in} y\equiv V(0)/A_1(0),
\end{equation}
and the $q^2$ dependence of the form factors. For the definitions of the form factors, see Bauer, Stech and Wirbel  \cite{ref:bsw1}. No particular model for the form factors was assumed in [3]. Instead, $R_h$ was studied as a function of $x$ and $y$ in three different scenarios for the $q^2$ dependence of the form factors.

The result of [3] was that the $(x,y)$ domain allowed by $R_h$ was inconsistent with the measurement of $x$ and $y$ from the semileptonic data in Ref. \cite{ref:kk}, and just barely consistent with that of Ref. \cite{ref:plf} . The allowed domain of $x$ and $y$ was also inconsistent  with the theoretical prediction of \cite{ref:bsw1}. Ref. \cite{ref:gour} also concluded that within the factorization scheme, the allowed range of $x$ and $y$ implied the following limits on the longitudinal polarization:
\begin{eqnarray}
\mbox{Monopole form factors with pole-mass 2.53 GeV:} \hspace{.5in}   0.43 \leq P_L \leq 0.55, \nonumber\\
 \indent \mbox{Monopole form factors with pole-mass 3.50 GeV:} \hspace{.5in}      0.33 \leq P_L \leq 0.55, \nonumber\\
 \indent \mbox{Flat form factors:} \hspace{2.0in}       0.36 \leq P_L \leq 0.55.
\end{eqnarray}

Subsequently, authors of \cite{ref:kasa} incorporated nonfactorized contributions in the decay matrix elements, and using the average of $x$ and $y$ from data \cite{{ref:kk},{ref:plf},{ref:pa2}}, showed that $R_h$ of (2) and
\begin{equation}
{R}_{s\ell} \equiv B(D_s^+ \rightarrow \phi {\ell}^{+} {\nu}_{\ell}) / B(D_s^+\rightarrow \phi {\pi}^{+}) = 0.54 \pm 0.10,~ \cite{ref:msw}
\end{equation}
could be understood within a scenario where the form factors have  a monopole dependence as in \cite{ref:bsw1}. However, there had to be significant nonfactorization contribution to $D_s^+ \rightarrow \phi {\pi}^{+}$, though factorization need not be violated in $D_s^+ \rightarrow \phi {\rho}^{+}$. Ref. \cite{ref:kasa} did not study longitudinal polarization.

An important point to be made is that there are three partial waves in $P\rightarrow VV$ decays, S, P and D, and though the decay rate does not depend on their phases, the longitudinal polarization does depend on the phase difference ${\delta}_{S} - {\delta}_{D}$. Ref. \cite{ref:gour} did not consider the effect of partial wave amplitude phases.

In this paper we have studied the data shown in (1) within the context of factorization invoking several form factor models (to be revealed in the next section), and allowing for nonzero S, P and D wave phases.
This paper is organized as follows: Section II deals with the details and the calculation.  A discussion of the results follows in Section III.

\begin{center}
{\bf II. Details of Calculation}
\end{center}

The decay $D_s^+ \longrightarrow {\rho}^+   \phi $  is Cabibbo-favored and is induced by the effective weak Hamiltonian given by
\begin{equation}
H =  \frac{G_{F}}{\sqrt{2}} V_{cs} V^{*}_{ud} [ C_{1}( \bar{u} d)( \bar{s} c)  +  C_{2} (\bar{u} c)( \bar{s} d) ],
%\label{eq:H} \\
\end{equation}
where $V_{qq'}$ are the CKM matrix elements; $C_1$and $C_2$ are the Wilson coefficients . The brackets $( \bar{u} d)$ represent $( V - A )$ color-singlet Dirac bilinears. Fierz transforming in color space with $N_c = 3$
\begin{equation}
(\bar{u} c)( \bar{s} d) = \frac{1}{3}( \bar{u} d)( \bar{s} c) +\frac{1}{2}\sum_{a=1}^{8}{(\bar{u}{\lambda}^{a}d)(\bar{s}{\lambda}^{a}c}),
\end{equation}
 the relevant part of the Hamiltonian can be written in the following form
\begin{equation}
H =  \frac{G_{F}}{\sqrt{2}} V_{cs} V^{*}_{ud} [ a_{1}( \bar{u} d)( \bar{s} c)  + {C}_{2} {O}_{8}],
\end{equation}
where $a_1 = C_1 + {C_2 \over 3} = 1.09 \pm 0.04$ \cite{ref:kasa} and $ O_8 = \frac{1}{2}\sum_{a=1}^{8}{(\bar{u}{\lambda}^{a}d)(\bar{s}{\lambda}^{a}c})$.  ${\lambda}^{a}$ are the Gell-mann matrices. In factorization approximation one neglects the contribution from the octet current part, ${O}_{8}$, and the  matrix element of the first term is written as a product of  two current matrix elements. It should be pointed out that there are no W-annihilation or W-exchange terms in ${{D}_{s}}^{+}\rightarrow {\rho}^{+} \phi$ decay. However, hair-pin graphs are allowed. We neglect them in what follows. The decay amplitude then  takes the following form:
\begin{equation}
A(D_s^+ \longrightarrow {\rho}^+   \phi) = {G_F \over \sqrt{2} } V_{cs}V_{ud}^*a_1\left\langle\phi \mid\bar{s} c\mid D_s^+ \right\rangle\left\langle \rho^+\mid \bar{u} d \mid 0\right\rangle.
\end{equation}
Each of the current matrix elements can be expressed in terms of meson decay constants and  invariant form factors. We use the following definitions:
\begin{eqnarray}
\left\langle \rho^+\mid \bar{u} d \mid 0\right\rangle &=&m_{\rho}f_{\rho}\varepsilon_{\mu}^* \\
\vspace{7mm}
<\phi \mid\bar{s} c\mid D_s> &=& {2 \over m_D + m_{\phi}} {\epsilon}_{\mu\nu\rho\sigma}\varepsilon_{\phi}^{*\nu}P_{D}^{\rho}P_{\phi}^{\sigma} V(q^2) +i\{\varepsilon_{\phi \mu}^*(m_D + m_{\phi})A_1(q^2) -  \nonumber\\
\vspace{7mm}
&& {\varepsilon_\phi^*.q \over m_D + m_\phi} (P_\phi + P_D)_\mu A_2(q^2) -{\varepsilon_\phi^*.q \over q^2} 2 m_\phi q_\mu A_3(q^2) + \nonumber\\
\vspace{7mm}
&&  {\varepsilon_\phi^*.q \over q_2} 2 m_\phi q_\mu A_0(q^2) \},
\end{eqnarray}
 where $ q = P_D -P_\phi$ is the momentum transfer, $f_{\rho}$ ( for which we use 212.0 MeV ) is the decay constant of the $\rho$ meson, $\varepsilon_{\phi (\rho)}$ is the polarization vector of the vector mesons $ \phi$( $ \rho$),  $ A_i(q^2), ( i = 1, 2, 3 )$ and $V(q^2)$ are invariant form factors defined in \cite{ref:bsw1} . The decay rate is given by
\begin{equation}
\Gamma(D_s^+ \rightarrow \rho^+\phi ) = {{\it p} \over 8\pi m_D^2}\left\{ |A_{00}|^2 + |A_{++}|^2 + |A_{--}|^2 \right\}
\end{equation}
where {\it p} is the centre of mass momentum in the final state. $A_{00}, A_{++} $ and $A_{--}$ are the longitudinal and transverse helicity amplitudes given by:
\begin{equation}
A_{00}(D_s^+ \rightarrow \rho^+\phi ) = - i {{G}_{F} \over \sqrt{2} }V_{cs}V_{ud}^{*}m_\rho f_\rho (m_D + m_{\phi}) a_1\left\{ a A_1(m_\rho^2) - bA_2(m^2_\rho) \right\}
\label{eq:a00}
\end{equation}
where the parameter $a$ and $b$ are defined as follow,
\begin{eqnarray}
 a &=& {1 -{r}^{2} - {t}^{2} \over 2rt}, ~~~~b = {k^2 \over 2rt(1+r)^2},~\mbox{ with} \nonumber \\
 r = {m_\phi \over m_D},~~ t = {m_\rho  \over m_D},~~  k^2 &=& (1 + r^4 + t^4 -2r^2 - 2t^2 - 2r^2t^2).
\end{eqnarray}
The other two helicity amplitudes are:
\begin{equation}
A_{\pm\pm}(D_s^+ \rightarrow \rho^+\phi ) =  i{{G}_{F} \over \sqrt{2} }V_{cs}V_{ud}^*m_\rho f_\rho (m_D + m_{\phi}) a_1\left\{ A_1(m_\rho^2) \pm {k \over (1 + r)^2}V(m^2_\rho) \right\}.
\label{eq:apm}
\end{equation}
The longitudinal polarization is defined by the ratio of the longitudinal decay rate to the total decay rate
\begin{equation}
{P}_{L} = {{\Gamma}_{00} \over \Gamma} = {\mid A_{00}\mid^2 \over \mid A_{++}\mid^2 + \mid A_{--}\mid^2 +\mid A_{00}\mid^2}.
\end{equation}
One can work with the helicity amplitudes or the partial wave amplitudes. We prefer to work with the latter as the dependence of the polarization on the partial wave phases is more obvious in that basis. The helicity and partial wave amplitudes are related by \cite{ref:nita} ,
\begin{equation}
A_{00} = -{1 \over \sqrt{3} }S +\sqrt{{2 \over 3}}D,~~~A_{\pm\pm} = {1 \over \sqrt{3}}S \pm {1 \over \sqrt{2} } P + {1 \over \sqrt{6}}D.
\label{eq:0pm}
\end{equation}
The partial waves are in general complex and can be expressed in terms of their phases as follow
\begin{equation}
S = \mid S\mid\exp{i\delta_S} ,~~P = \mid P\mid \exp{i\delta_P},~~ D = \mid D\mid \exp{i\delta_D}.
\label{eq:spd}
\end{equation}
The decay rate is given by an incoherent sum,  $\Gamma \propto  \mid A_{++}\mid^2 + \mid A_{--}\mid^2 + \mid A_{00}\mid^2  = \mid S\mid^2 + \mid P\mid^2 + \mid D\mid^2$, and is independent of the phases.  But the polarization does depend on the phase difference $\delta_{SD} = \delta_S - \delta_D$ arising from the interference between S and D waves
\begin{equation}
P_L = {1 \over 3} {\mid S\mid ^2 + 2 \mid D\mid^2 - {2\sqrt{2} }\mid S\mid\mid D\mid \cos{\delta_{SD}} \over \mid S\mid^2 + \mid P\mid^2 + \mid D\mid^2}.
\end{equation}

S, P and D waves were calculated first by using the amplitudes (\ref{eq:a00}) and (\ref{eq:apm}) in (\ref{eq:0pm}) and feeding in the phases by hand as shown in (\ref{eq:spd}). To continue with the numerical analysis of the decay rate $\Gamma$ and the longitudinal polarization $P_L$,  we have used form factors from six different sources: i) Bauer, Stech and Wirbel (BSW$I$) \cite{ref:bsw1}, where an infinite momentum frame is used to calculate the form factors at $q^2 = 0$, and a monopole form ( pole masses are as in \cite{ref:bsw1} ) for $q^2$ dependence is assumed to extrapolate all the form factors to the desired value of $q^2$; ii) BSW$II$ is a modification of BSW$I$, where while $F_0(q^2)$ and $A_1(q^2)$ are the same as in BSW$I$, a dipole $q^2$ dependence is assumed for $A_2(q^2)$ and $V(q^2)$; iii) Altomari and Wolfenstein (AW) model \cite{ref:aw}, where the form factors are evaluated in the limit of zero  recoil, and a monopole form is used to extrapolate to the desired value of $q^2$; iv) CDDGFN model \cite{ref:cdd}, where the form factors are evaluated at $q^2 = 0$ in an effective Lagrangian satisfying heavy quark spin-flavor symmetry in which light vector particles are introduced as gauge particles in a broken chiral symmetry.  A monopole form is used for the $q^2$ dependence ( we mention here that we have updated this model by using more recent experimental results of the form factors $A_1^{DK^*}(0)$,  $A_2^{DK^*}(0)$and  $V^{DK^*}(0)$ \cite{ref:pdg}, and $f_{D_s} = 241 \pm 37$ MeV \cite{fer:rich} in calculating the weak couplings constant, of the model, at $q^2 = 0$ \cite{ref:cdd} , which are subsequently used in evaluating the required form factors ); v) Isgur, Scora, Grinstein and Wise (ISGW) model \cite{ref:isgw}, where a non-relativistic quark model is used to calculate the form factors at zero recoil and an exponential $q^2$ dependence is used to extrapolate them to the desired $q^2$; vi) using experimental values \cite{ref:pdg} of the form factors at $q^2 = 0$ and extrapolating them using monopole forms.

\begin{center}
{\bf III. Results and Discussion}
\end{center}

The results are summarized in Table 1. For the entries in the last column of Table 1 we have used the experimental values of the form factors at $q^2 = 0$: $A_1^{D_S\phi}(0) = 0.62 \pm 0.06, A_2^{D_S\phi}(0) = 1.0 \pm 0.3, V^{D_S\phi}(0) = 0.9 \pm 0.3$ \cite{ref:pdg} and extrapolated them with monopole forms. First, we note from the Table that all models, except the CDDGFN and the one where experimentally measured form factors are used, overestimate the decay rate. This fact arises from an overestimate of the form factor $A_{1}$. Ref.\cite{ref:bc} has noted this fact and attributes it to the imposing of chiral symmetry. Further, as Ref. \cite{ref:cdd1} has argued, more theoretical as well as experimental studies are needed for a better understanding of the $q^2$ dependence of  form factors. Second, we observe that all the six sources of form factors allow a range for the polarization which overlaps with experiment with $\delta_{SD} \neq 0$. Note that the polarization is independent of the normalization of $A_{1}$. It is also found that most of the final state in the decay  $D_s^+ \rightarrow \rho^+\phi $ is in the $S$ wave. It is also seen from the Table that the hierarchy of the partial wave amplitudes is: $ \mid S \mid > \mid P \mid > \mid D \mid$. If we consider the final state to get contribution only from $S$ wave the decay rates would only be reduced by (5 to 12) \%, while the polarization would be $P_L = 0.33$, The hierarchy of the sizes of the partial wave amplitudes is in accordance with intuitive expectations based on threshold arguments. It is the S-wave dominance which makes an accurate determination of $\delta_{SD}$ difficult (the errors in $\delta_{SD}$ are large despite the fact that the errors in $P_{L}$ are small) since the D-wave is an order of magnitude smaller than the S-wave. The interference term is, consequently, small.

 %------------------------------------------------------------------------------------------------------
\begin{table}
\centering
\caption{Decay rate and longitudinal polarization for $D_s^+ \longrightarrow {\rho}^+   \phi $.  The values of $\Gamma$ must be multiplied by $10^{12}s^{-1}$. $\delta_{SD} = \delta_{S} - \delta_{D}$ is the value needed to get agreement with $P_{L}$ data to one STD. The last column uses experimentally measured form factors. 'Expt.FF' stands for 'Experimental form factors'.}
\vspace{7mm}
\begin{tabular}{|c|c|c|c|c|c|c|c|} \hline
$$&$BSWI$&$BSWII$&$AW$&$CDDGFN$&$ISGW$&$Expt.FF$ \\ \hline
$\Gamma$&$0.32$&$0.32$&$0.35$&$0.15$&$0.37 $&$0.18\pm0.04 \cite{ref:pdg}$ \\
$\delta_{SD}$&$135 \pm 45 $&$138 \pm 43 $&$122 \pm 32$&$140 \pm 40 $&$120 \pm 35 $&$134 \pm 46 $ \\
${|S| \over |P|}$&$4.3$&$3.7$&$3.8$&$2.8$&$5.5$&$4.7$\\
${|S| \over |D|}$&$11.9$&$13.5$&$7.4$&$8.2$&$8.6$&$16.5$\\
\hline
\end{tabular}
\begin{tabular}{|c|c|} \hline
$\mbox{ Experimental values of $\Gamma$ and $P_L$}$&$\Gamma =0.14\pm0.05$ \cite{ref:pdg},$P_L = 0.370\pm 0.052$\cite{ref:balic}\\
\hline
\end{tabular}

\label{tab:DCSISO}
\end{table}
\vspace{0.25cm}
{\bf Aknowledgments}:
The author would like to thank Dr. A. N. Kamal for valuable discussions during the course of this work and a thorough reading of this article. This research was partially funded by the Natural Sciences and Engineering Research Council of Canada through a grant to Dr. A. N. Kamal.

\pagebreak

\end{document}